\documentclass{sf2a-conf2012}
\usepackage{graphicx}
\usepackage{hyperref}
\usepackage[]{natbib}  
\usepackage[]{aeguill}
\usepackage{epstopdf}

\def\BibTeX{{\rm B\kern-.05em{\sc i\kern-.025em b}\kern-.08em
    T\kern-.1667em\lower.7ex\hbox{E}\kern-.125emX}}
\bibpunct{(}{)}{;}{a}{}{,}  

\begin{document}

\TitreGlobal{SF2A 2012}


\title{Simulated histories of reionization with merger tree of HII regions}

\runningtitle{Merger tree of HII regions}

\author{Jonathan Chardin}\address{Observatoire Astronomique de Strasbourg, Universit\'e de Strasbourg, CNRS UMR 7550, 11 rue de l'Universit\'e, F-67000 Strasbourg, France}

\author{Dominique Aubert$^1$}

\setcounter{page}{237}

\index{J.C}
\index{D.A.}


\maketitle


\begin{abstract}

We describe a new methodology to analyze the reionization process in numerical simulations:
The evolution of the reionization is investigated by focusing on the merger histories of individual HII regions.
From the merger tree of ionized patches, one can track the individual
evolution of the regions such as e.g. their size, or investigate the properties 
of the percolation process by looking at the formation rate, 
the frequency of mergers and the number of individual HII regions involved in the mergers.
By applying this technique to cosmological simulations with radiative
transfer, we show how this methodology is a good candidate to quantify the
impact of the star formation adopted on the history of the reionization. As
an application we show how different models of sources result in different
evolutions and geometry of the reionization even though they produce
e.g. similar ionized fraction or optical depth.

\end{abstract}

\begin{keywords}
Reionization, HII regions , first stars - Methods: numerical
\end{keywords}


\section{Introduction}

The reionization of the Universe occurred between $z \sim 20$ and $z \sim 6$.
During this period the first generation of ionizing stars created a multitude of HII regions.
Therefore a challenge was to simulate the reionization process numerically (see \citealt{2009arXiv0906.4348T} for a complete review of these models).
In this context, multiple approach can be overtaken in order to investigate the reionization in numerical simulation.
Usually the time sequence of the process is explored by focusing on global quantities such as the evolution of the averaged ionized fraction
or the optical depth evolution (\citealt{2006MNRAS.369.1625I} and \citealt{2010ApJ...724..244A} for example).

We propose here to present an alternative method that allows us to see the reionization through the evolution of the HII regions thanks to a merger tree.
Such a merger tree enables us to follow the evolution of the individual HII regions properties and then leads to a `local' perspective with multiple histories
of reionization instead of one `general' scenario with the quantities commonly used to analyze the simulations.

We apply here this methodology to three simulations of reionization. In each of them we vary the prescription to generate the ionizing sources.
We thus aim to show in what the merger tree approach is a good estimator to quantify the differences induced by the ionizing source models in the related histories of reionization.

\section{Methodology}

\subsection{\textit{friend-of-friend} algorithm}

Firstly we need to identify the individual HII regions in each snapshots of the simulation. 
We have assumed that a cell of the grid is ionized if its ionization fraction $x \ge 0.5$.
We then explore the box and when we encounter a ionized cell we allocate to it an 
identification number corresponding to the ionized region being tested. 
Then, the \textit{friend-of-friend} algorithm proceeds by allocating to the ionized nearest neighbors of this cell the identification 
number of the HII regions being explored.
We are therefore able to separate all the ionized regions and to keep track of each of them with the identification number.

\subsection{Merger tree}

Secondly we can build the merger tree itself in order to follow with time the evolution of the properties of the HII regions. 
We simply proceed by following the evolution of the identification numbers of the HII regions allocated during the identification step. 
In practice, we extract where are located the cells of an HII region at time $t$ and 
look at the identification number that they received at time $t+1$. We then link the two identification number
between the two snapshots. We repeat this process for all the HII regions and between all the snapshots of the simulation.

\begin{figure*}
 \begin{center}
    \begin{tabular}{cc}
      \includegraphics[width=6cm,height=5cm]{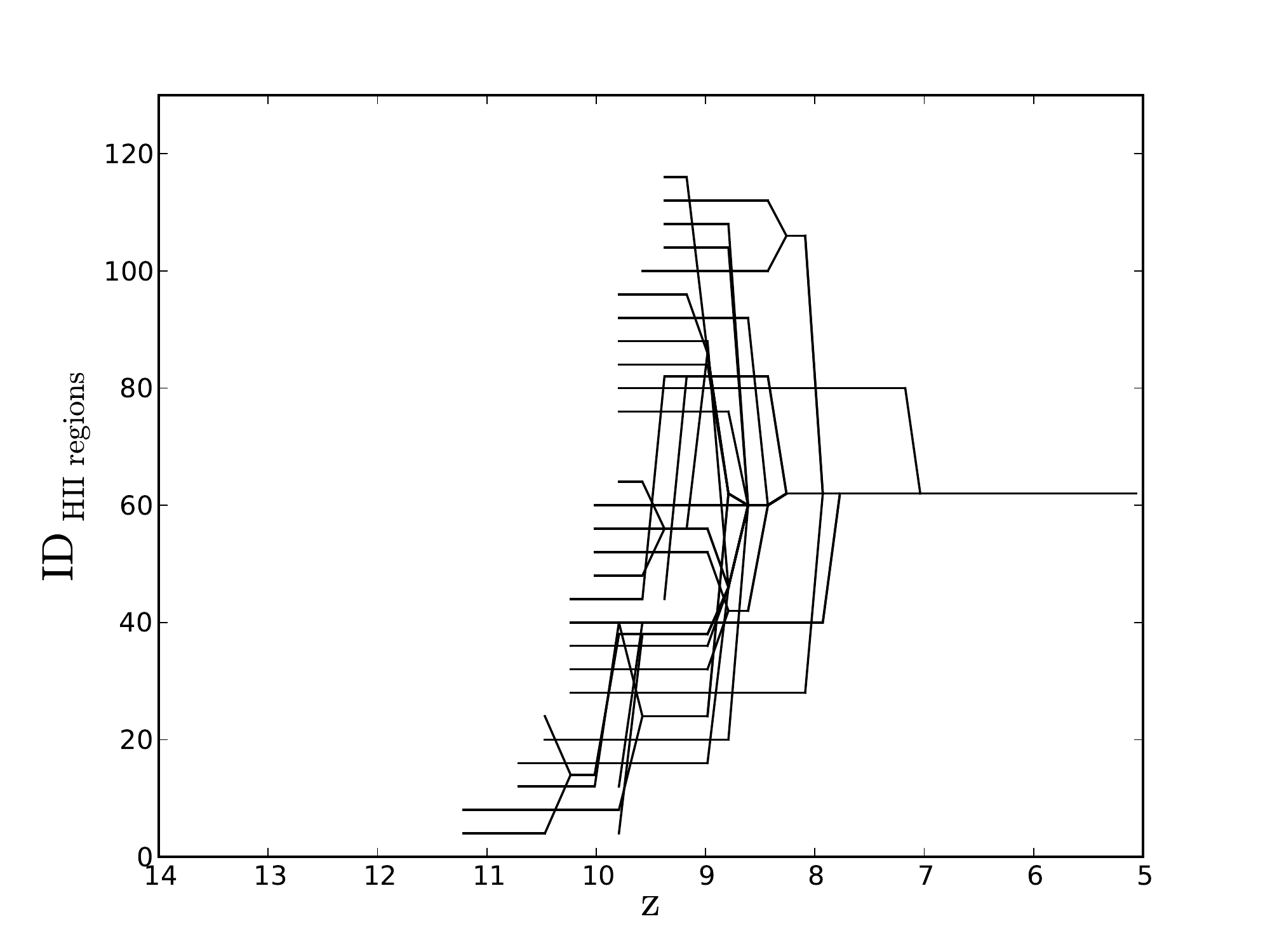} &
      \includegraphics[width=6cm,height=5cm]{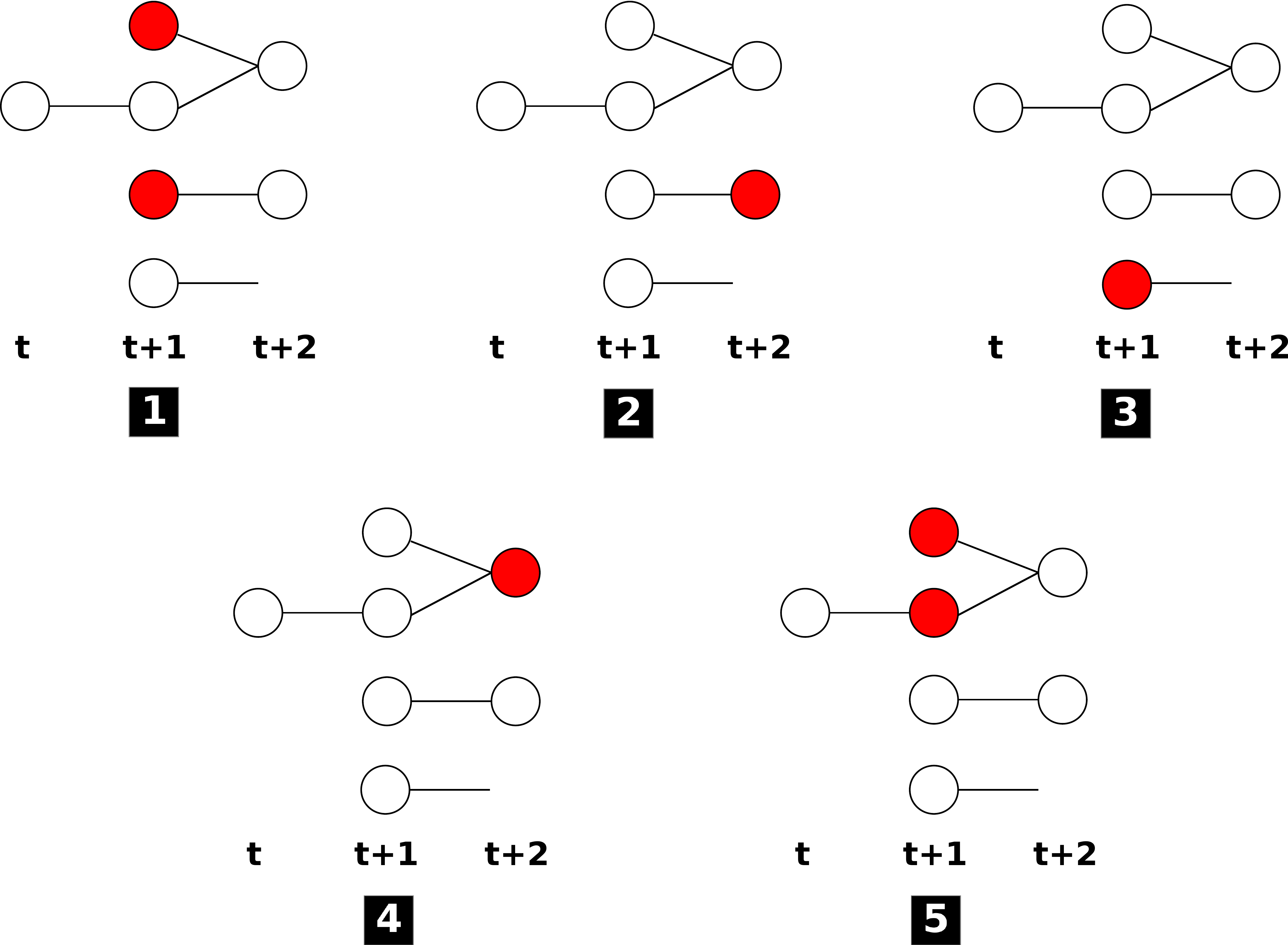} \\
       (a) &   (b) \\
    \end{tabular} 
\caption{Panel (a): An illustration of the merger tree of HII regions. 
Each black line represents an ID evolution with the redshift for a distinct HII region.
For clarity, we represent here only the ID evolution for 30 regions.
Panel (b): representation of the typical properties investigated with the merger tree. In each diagram,
red items symbolize the kind of properties that we follow. 1: the number of new HII regions between two snapshots, 
2: the number of growing ionized regions, 
3: the number of HII regions which recombine,
4: the number of HII regions resulting from mergers
and, 5: the number of parents involved for an HII region resulting from mergers. }
\label{a_view_of_the_merger_tree}
\end{center}
\end{figure*}

Panel (a) of the figure \ref{a_view_of_the_merger_tree} shows a representation of the merger tree for 30 regions.
The (b) panel shows the typical properties that can be investigated with the merger tree.

\subsection{simulations}

We propose in this work to apply the merger tree methodology on three simulations of cosmic reionization to compare the 
impact of three different ionizing source models on the observed reionization history.
We performed two simulation with stellar particles generated with the RAMSES code (\citealt{2002A&A...385..337T}).
These two models are based on the same sources maps and only differ in regard to the adopted emissivity law with time for the sources.
We also performed a third model where the dark matter halos extracted from the density fields are assumed as ionizing sources with an emissivity
proportional to the halo masses.
In all cases the radiative transfer is done with the ATON code (\citealt{2008MNRAS.387..295A}).
The simulations properties of each model are reported in table \ref{simus_table}.

\begin{table}[]
\begin{center}
  \begin{tabular}{|c|c|c|c|}
  \hline {\bf Model name}               & {\bf Box size}              & {\bf Source type} &  {\bf Emissivity} \\
  \hline {Boosted Star}               & {200 Mpc/h}              & {Stellar particles} &  {Converged number of photons at each instant} \\
       { }               & { }              & { } &  {(Decreasing emissivity with time)} \\
  \hline  {Star}                      & {200 Mpc/h}              & {Stellar particles} &  {Converged number of photons at $z \sim 6$} \\
       {}                      & {}              & {} &  {(Constant emissivity with time)} \\
  \hline	  {Halo}                     & {200 Mpc/h}              & {DM halo} &  {Proportional to halo masses} \\
  \hline
  \end{tabular}
  \caption{Summary of the simulations properties.}
 \label{simus_table}
\end{center}
\end{table}

\section{Results}

\subsection{Number of HII regions}

\begin{figure*}
   \begin{center}
    \begin{tabular}{ccc}
      \hskip -1.25truecm
      \includegraphics[width=6cm,height=5cm]{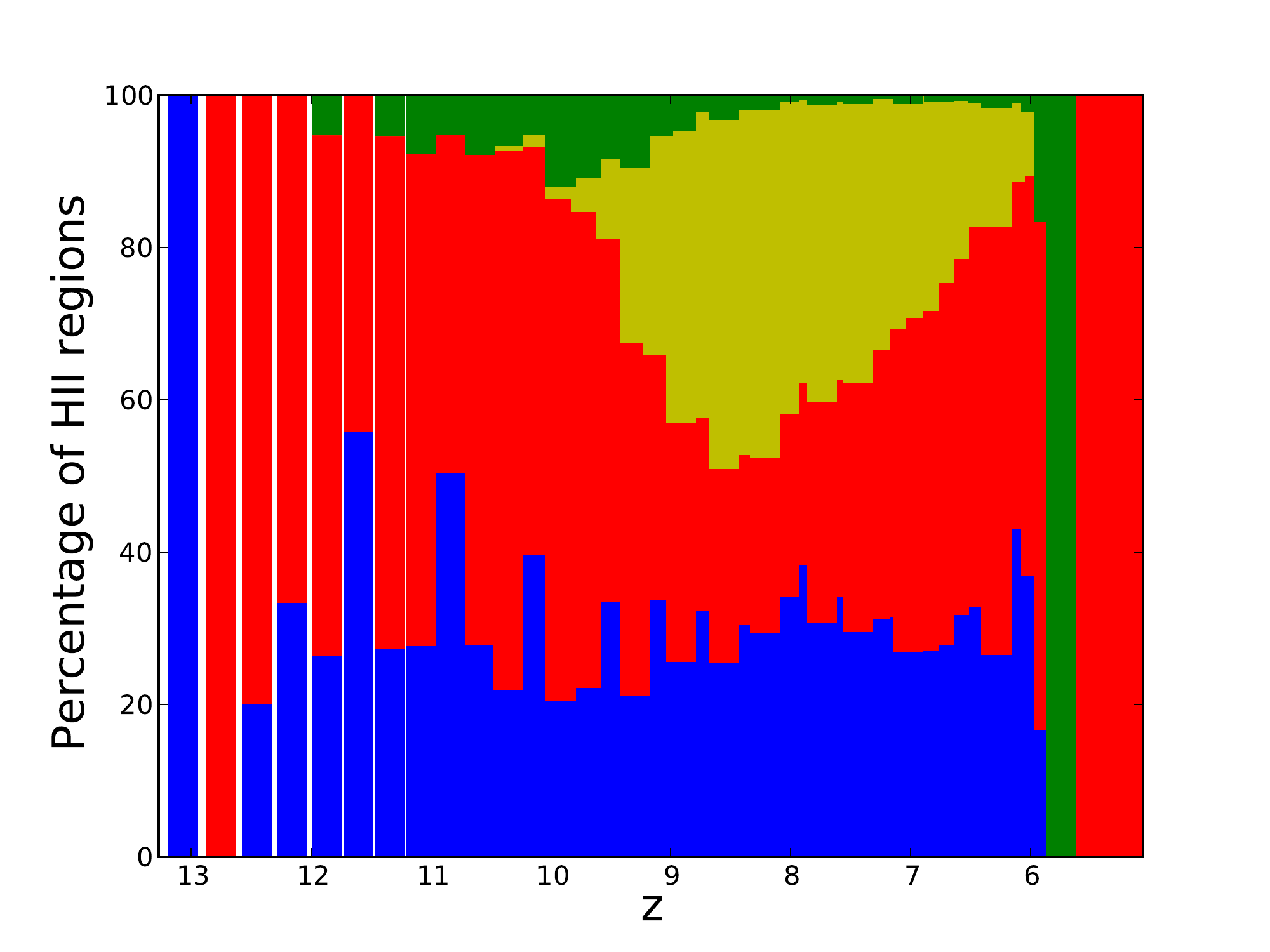} &
	\includegraphics[width=6cm,height=5cm]{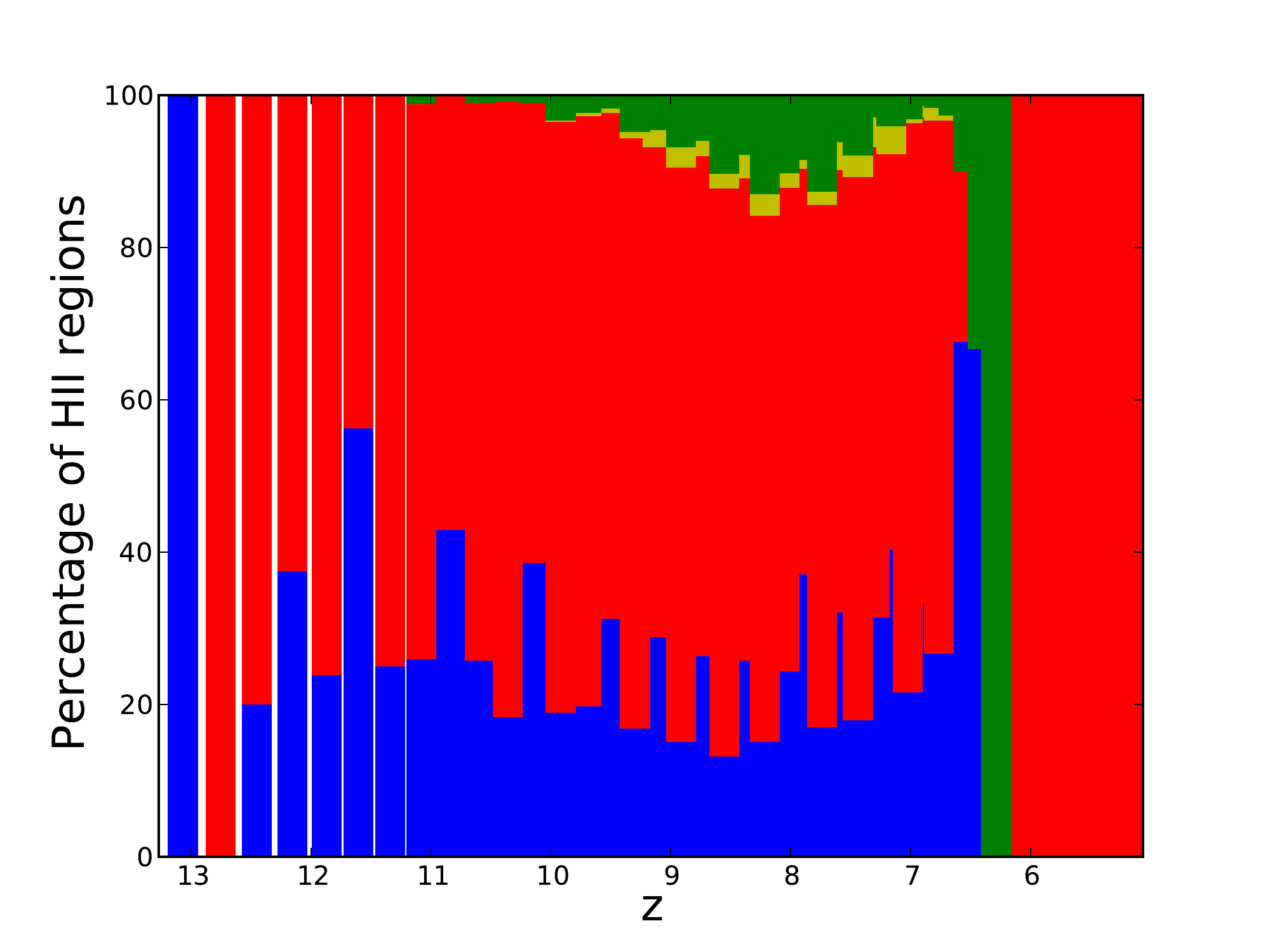} &
	\includegraphics[width=6cm,height=5cm]{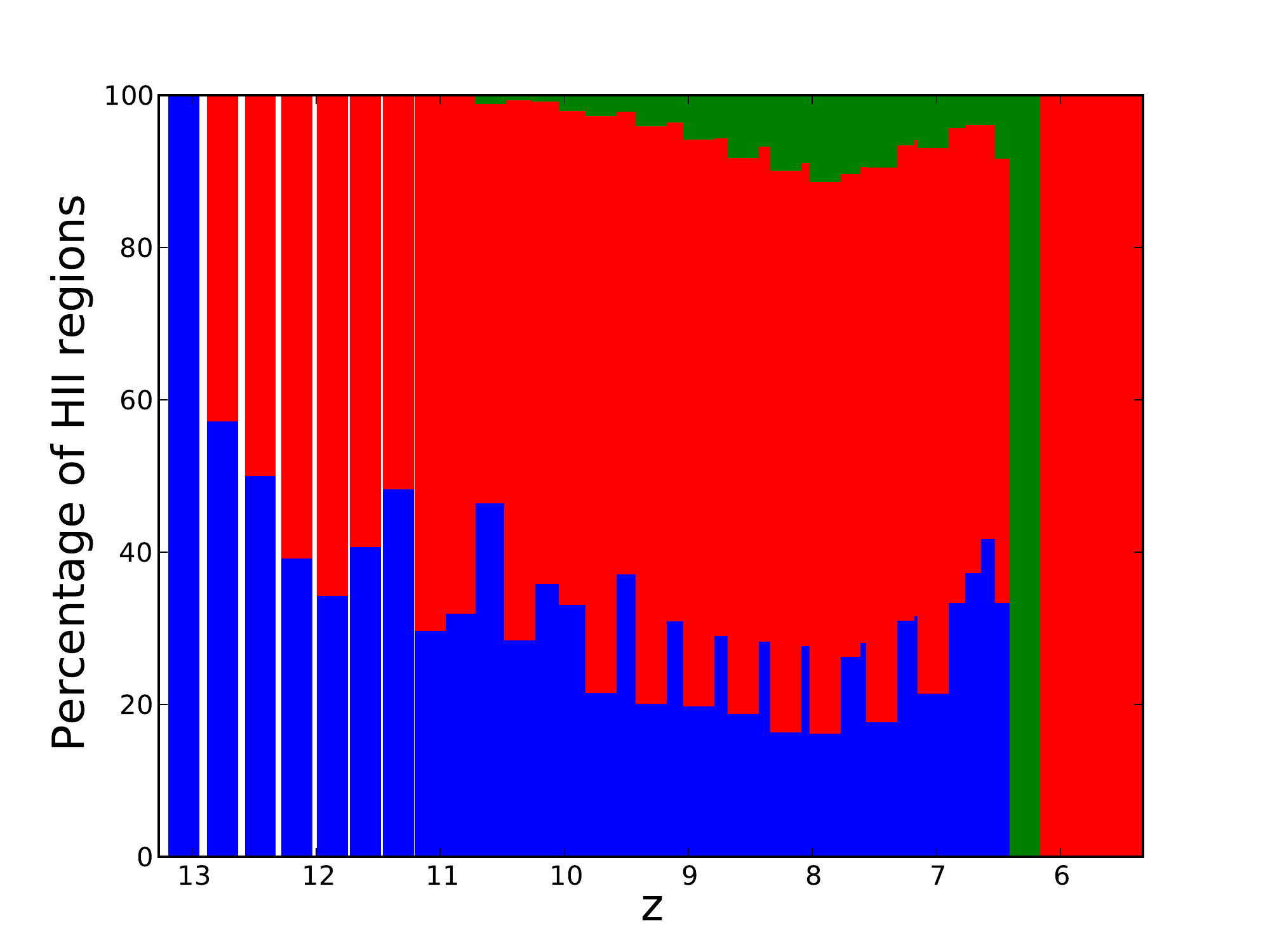} \\
     Boosted Star & Star & Halo \\
\end{tabular}    
  \caption{Evolution of the proportion of each kind of HII regions as a function of redshift for the three models of ionizing sources. 
The colors stands the new HII regions (blue), the expanding regions (red), 
the regions that will recombine (yellow) and, the regions resulting from mergers (green). The black vertical line shows the peak 
of the absolute number of HII regions: $\mathrm{z_{\,peak}}$.}
    \label{proportion}
  \end{center}
 \end{figure*}

Figure \ref{proportion} shows the evolution of the relative proportion of each kind of HII regions as a function of redshift.
The Star and Halo models present a significant proportion of regions resulting from merger around a redshift of $z \sim 8$.
On the other hand, this period where a significant population of regions resulting from mergers is detected happens earlier in the Boosted Star
model $9 \le z \le 12$.
Early large regions, that are created in early reionization due to the greater correction in the photons emissivity at high redshift, merge quickly in this model
and would lead to a main region that then drive the reionization process.
Conversely the HII regions in the Star and Halo models are smaller in early reionization and the individual growth process of HII region can be tracked during a longer period before
the \textit{overlap period}.  

On the other hand the Star model presents few recombination episodes from $z \sim 10$ while the Halo model presents no recombining regions detected.
Alternately the Boosted Star model shows a significant fraction of regions that recombine from $z \sim 10$.
In this last model these regions are the result of the dislocation of the early large HII regions that can not be sustained because the renewal 
rate of sources inside these regions, combined to their decreasing emissivities, is not powerful enough.
In the Star model the emissivity of sources is constant during the whole simulation. Thus, we find less recombining regions than in the Boosted Star 
model because recombinations only occur inside pre-existing regions where the renewal rate of sources is not sufficient.
Finally the Halo model shows no recombinations because the sources are more numerous than in both other models and have no finite lifetime.
Thus the regions have always smaller sizes and are always sustained compared to both other models.

\label{proportion_HII_regions}

\subsection{Sizes of HII regions}

\begin{figure*}
   \begin{center}
    \begin{tabular}{ccc}
      \hskip -1.25truecm
      \includegraphics[width=6cm,height=5cm]{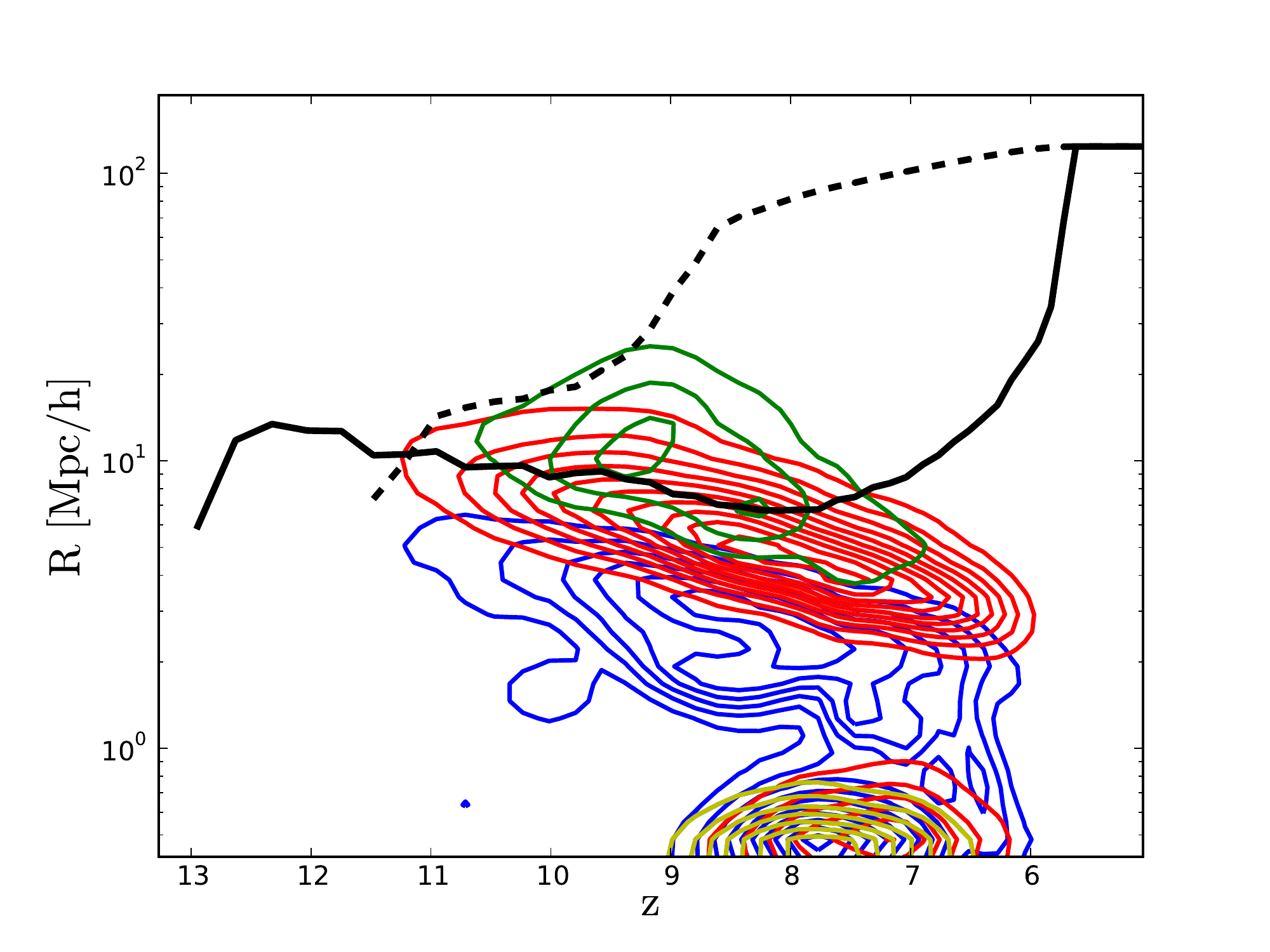} &
      \includegraphics[width=6cm,height=5cm]{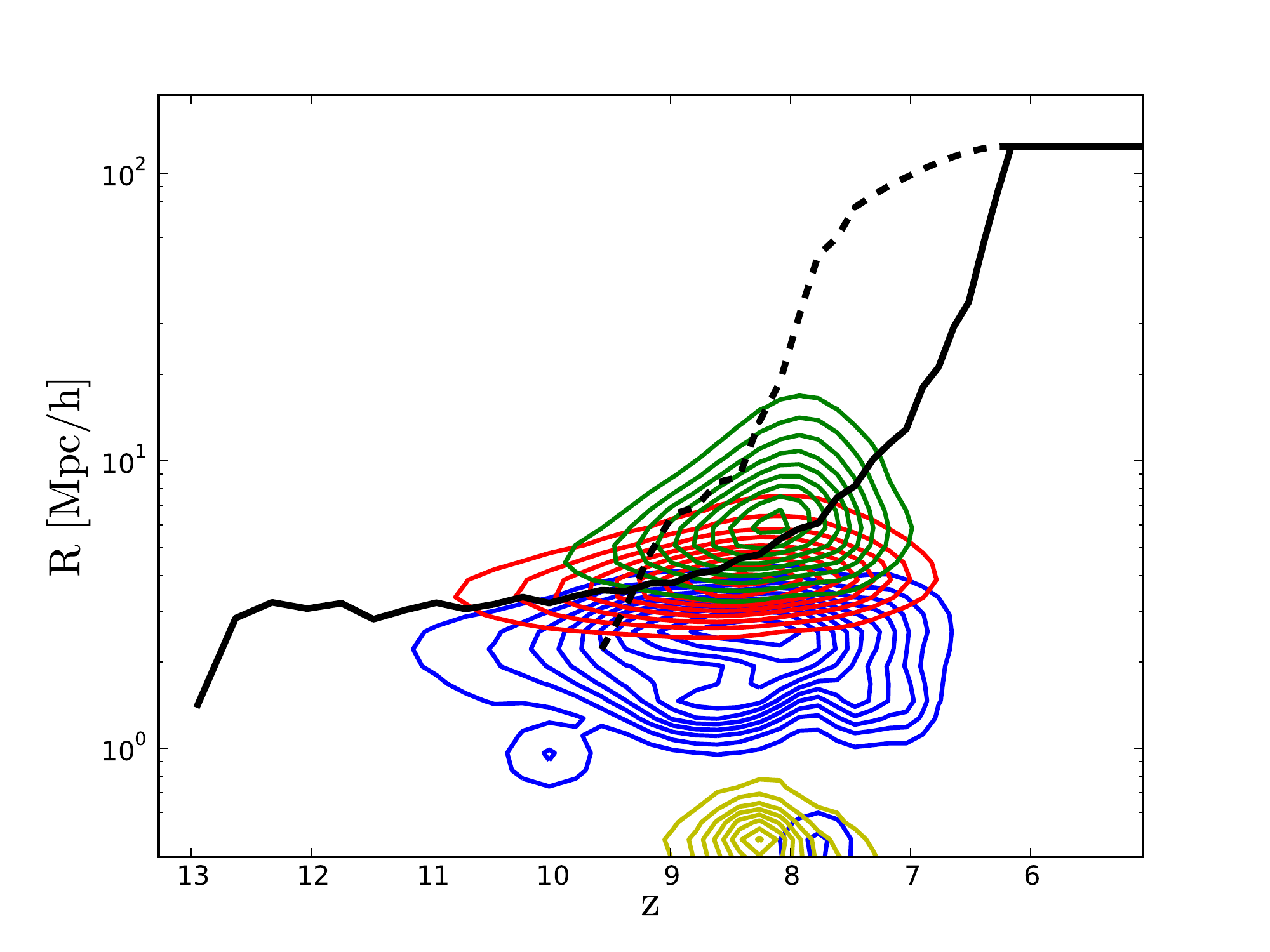}&
      \includegraphics[width=6cm,height=5cm]{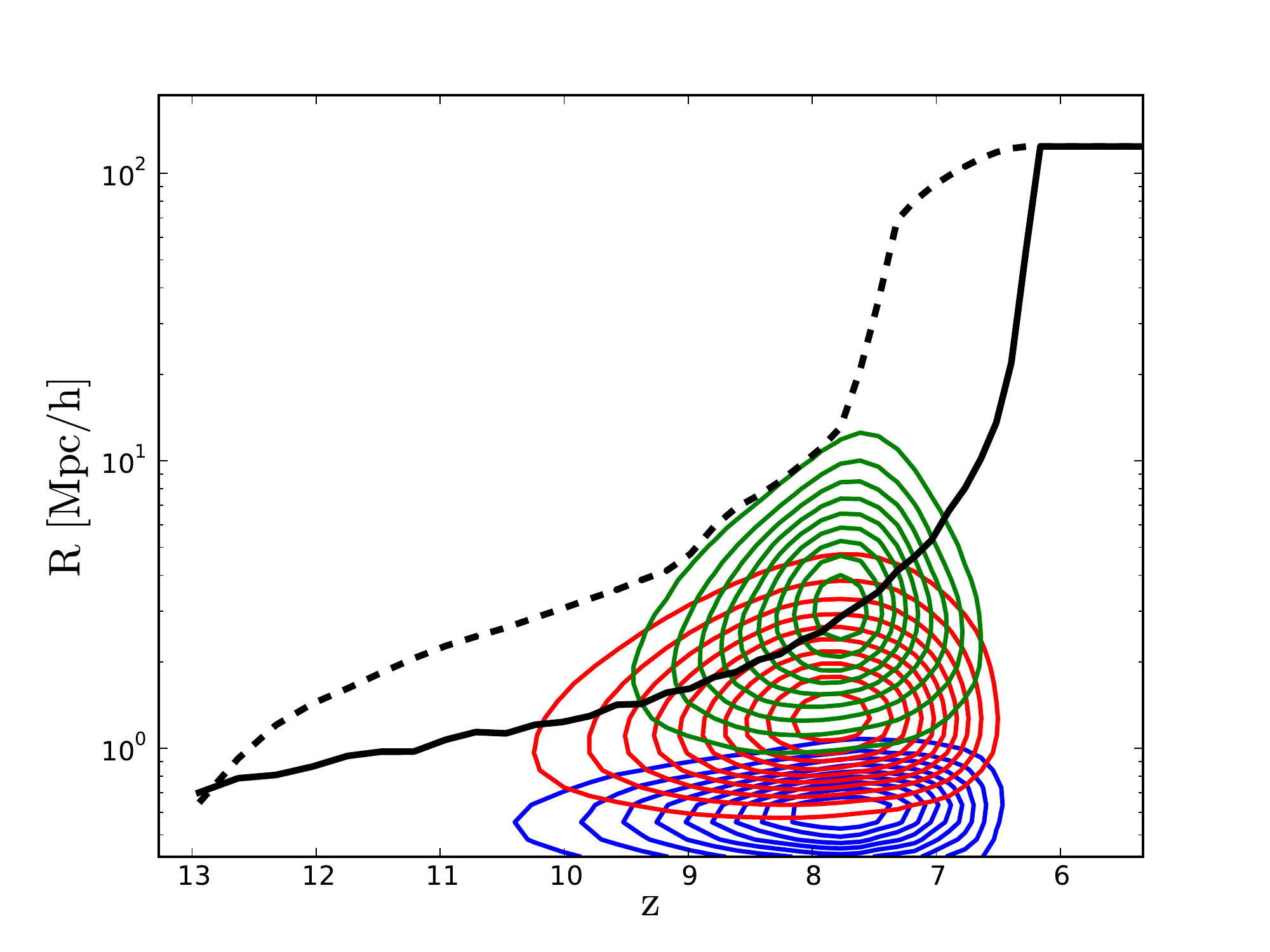} \\
    Boosted Star & Star & Halo \\
\end{tabular}    
   \caption{ Evolution of the radius distribution of each kind of HII regions as a function of redshift for the three models of ionizing source formation. 
The colors stands the new HII regions (blue), the expanding regions (red), 
the regions that will recombine (yellow) and, the regions resulting from mergers (green).}
    \label{isocontour_rayons}
  \end{center}
 \end{figure*}

Figure \ref{isocontour_rayons} shows the evolution of the radius distribution for the different kinds of HII regions with redshift according to the related color code.
The black solid line presents the evolution of the average radius for the detected regions and the black dashed line presents the evolution of the single last region
detected when the reionization is achieved. With the help of the merger tree we follow back in time this region and, each time, we calculate the radius of its main progenitor.

First, wee find that every kind of regions occupy a dedicated range of radius in the distribution in every model.
The new regions occupy the bottom of the distribution while the regions resulting from mergers are in the top and the expanding ones in the middle of the distribution.
We can note that the range of radii covered by a distinct type of region eventually overlap with the distribution of the other kinds of regions.
This reflects the fact that the time sampling of the simulation allows us to detect regions with same radii but belonging to different kinds of regions.

Second, we find in every models, that the evolution of the radius distribution traces the underlying law of evolution of the source emissivities.
Thus in the Star model we see an average constant radius for the HII region, at least during the pre-overlap period, that traces the constant boost emissivity of ionizing sources.  
The Boosted Star model shows a decreasing gradient for the radii of HII regions that typically reflects the decreasing emissivity of ionizing sources as reionization progresses.
Finally the Halo model presents the largest ranges of radii covered by each type of regions which is representative of the underlying range of halo masses that dictates the emissivity law
of ionizing sources in this case.

As expected before, we see from $z \sim 9$ that the recombining regions are the smallest regions in the distribution in the Boosted Star and the Star model.
This comfort us to say that these regions are small regions resulting from the dislocation of early large regions because they are detected as small patches around large HII regions
in fragmentation.

Finally we find the emergence of a single main region in size in every models followed with the black dashed line. 
This region appears at first in the Boosted Star model at $z \sim 9$ and would be typically the result of the rapid mergers of early large HII regions.
The moment of emergence of this region is delayed in the Star model at $z \sim 8$ and at $z \sim 7.5$ in the Halo model because the region are smaller and their individual
growing step can be tracked much longer than in the Boosted Star model.

\subsection{Merger of HII regions}

\begin{figure*}
   \begin{center}
    \begin{tabular}{ccc}
      \hskip -1.25truecm
      \includegraphics[width=6cm,height=5cm]{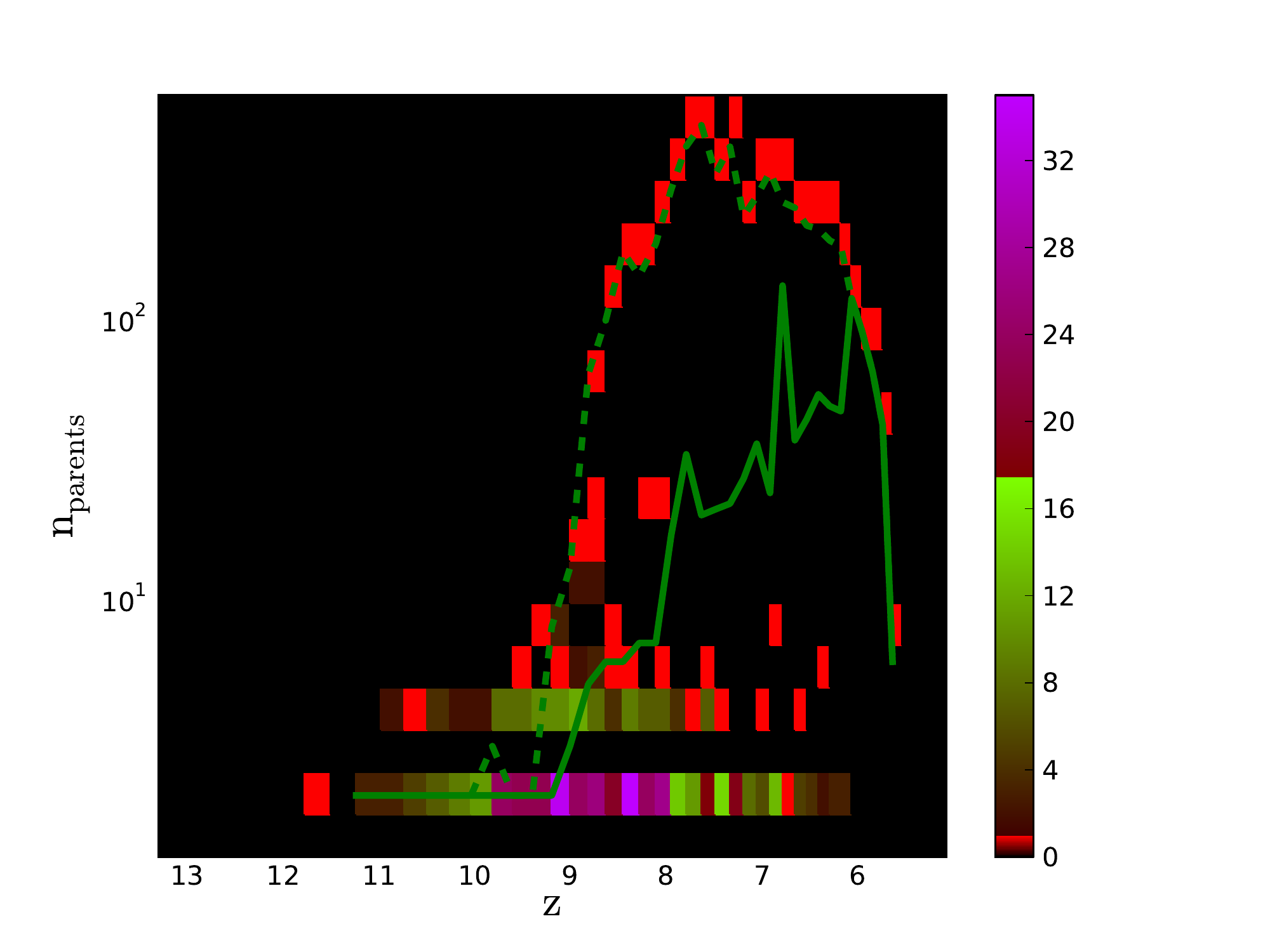} &
      \includegraphics[width=6cm,height=5cm]{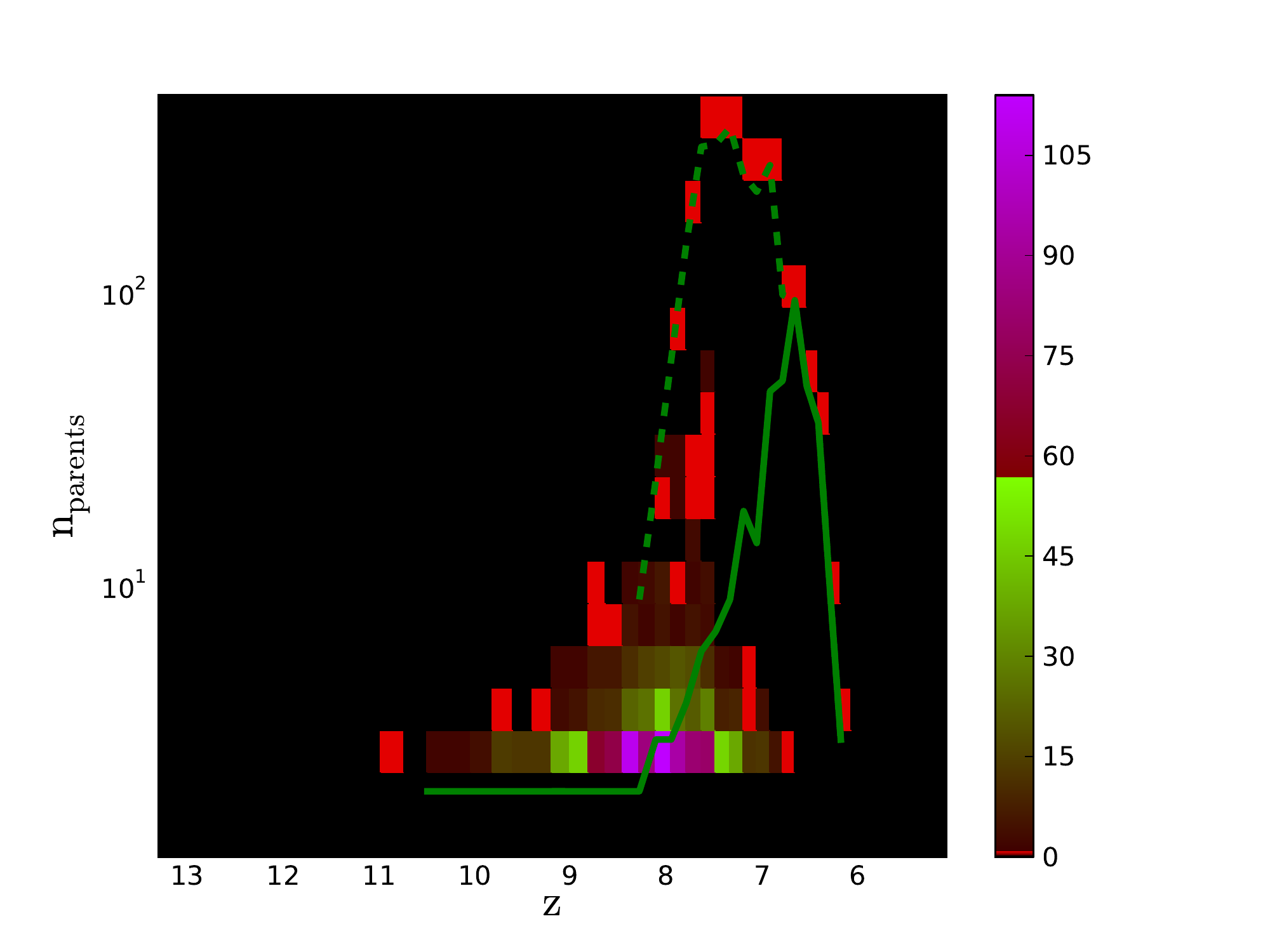}&
      \includegraphics[width=6cm,height=5cm]{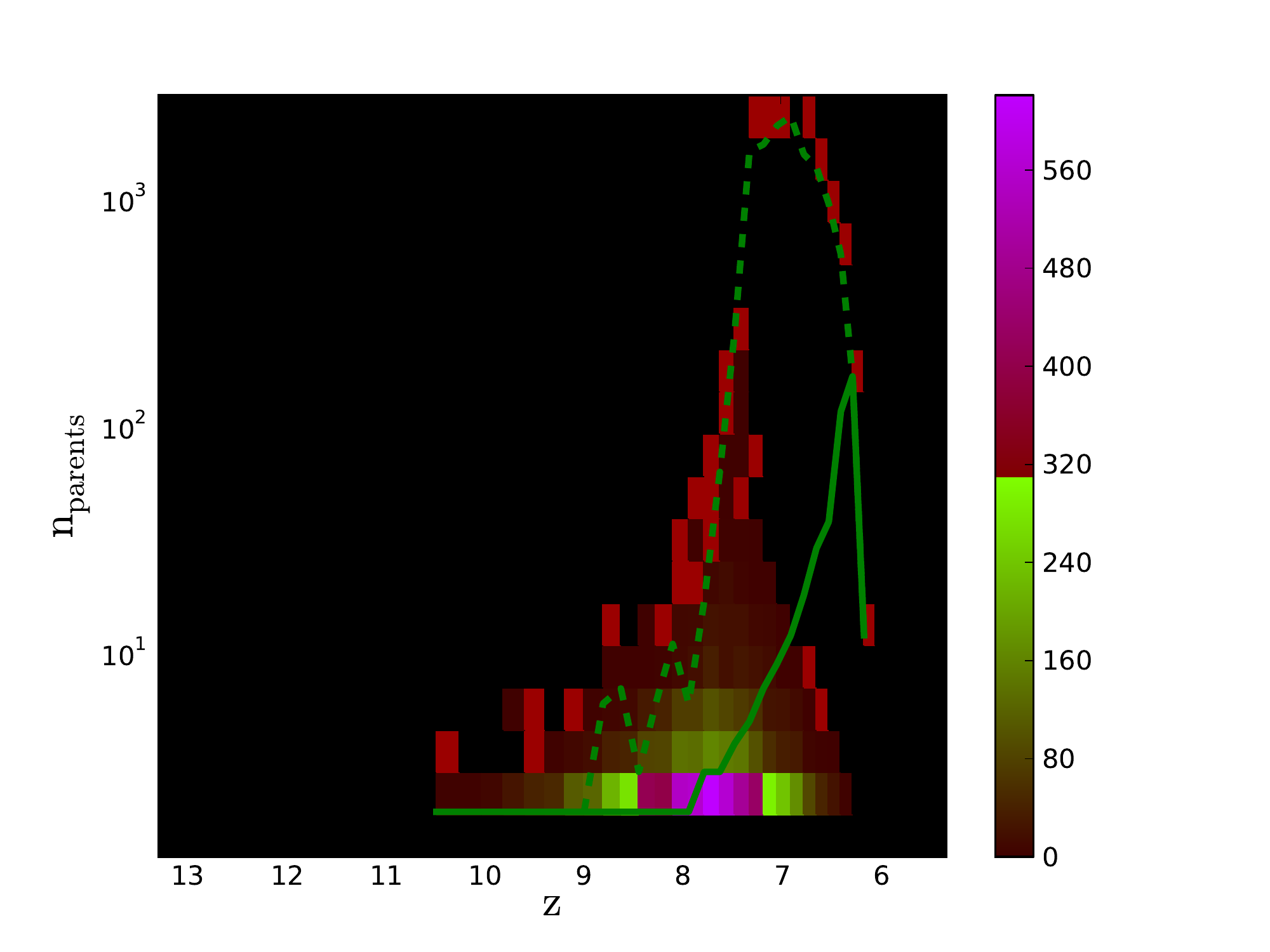} \\
    Boosted Star & Star & Halo \\
\end{tabular}    
   \caption{Distribution of the number of parents for the HII regions resulting from merger as a function of redshift for the three kinds of ionizing sources.
We show the average number of parents (solid green line) and the evolution of the number of parents of the main region in size (green dashed line).}
    \label{parents}
  \end{center}
 \end{figure*}

Figure \ref{parents} shows the evolution of the distribution of the number of parents for HII regions resulting from mergers.
Again, we represent the evolution of the average number of parents (solid green line) and the evolution of the number of parents of the main region in size (green dashed line).

First, we find, in every models, that the process of mergers between the HII regions occurs in a binary-tertiary manner as seen with 
the peak at 2-3 in the distributions during the whole simulated period.
This tells us that the time sampling of the simulations allows us to detect the individual mergers.

Second, in all models, we find the emergence of a region that concentrates the mergers with a greater number of parents in the distributions.
This region corresponds each time to the main region in size detected before. 
Indeed, the evolution of the  number of parents of this main region in size matches the evolution of the number of parents of this region that concentrates 
the mergers in the distributions.

This region appears, as already seen before, at first in the Boosted Star model at $z \sim 9$, then in the Star model at $z \sim 8$ and finally in the Halo model at $z \sim 7.5$.
The Star model shows few regions, except the main regions, that can reach a number of parents greater than 10 while the distribution in the Halo model is much smoother with a significant 
proportion of regions that can reach this number. In other words, the Halo model can track much longer the individual merger histories of multiple regions at the expense of a 
single region that impose its domination by phagocytising the others. 
The Boosted model finally shows only the main region that can reach a number of parents greater than 10. This is thus definitely in this model where the reionization is
driven by a single region that concentrates the most the mergers. This is finally the model the least able to track the individual merger episodes during a longer period of time
and thus the least able to follow the `local' reionization.

\section{Conclusions}

We have developed a new technique to analyze simulations of cosmic reionization.
By constructing a merger tree of HII regions and applying this measuring probe to three simulations, we
have shown how we can quantify the impact of the variation of the ionizing source models on the related histories of reionization 
(see \citealt{2012arXiv1210.1445C} for a more quantitative and detailed work on this topic).
We have demonstrated that a semi-analytic prescription for the ionizing sources based on dark matter halos is the most able to track the `local' histories of reionization.
We have also shown that we can match a similar history with the Star model that use stellar particles as ionizing sources with a constant emissivity during the whole experiment.
On the other hand the Boosted Star model, that uses the same source maps as the Star model but with a different emissivity evolution for the sources, presents a reionization history that sensibly
differ from both other models. This model presents an early large ionized region that manages the reionization merger history at the expense of the tracking of `local' evolutions.
We finally plan in the future to apply repetitively this technique to study the impact of the inputs in simulations of reionization.

\begin{acknowledgements}
We want to thank B.Semelin, P Ocvirk, R. Teyssier and H. Wozniak for comments and discussion.
This work is supported by the LIDAU ANR.
\end{acknowledgements}

\bibliographystyle{aa}  
\bibliography{biblio} 

\end{document}